\documentclass[prl,preprint]{revtex4}%
\usepackage{amsfonts}%
\usepackage{amsmath}%
\usepackage{amssymb}%
\usepackage{graphicx}

\begin{document}
\title{Phase Reversal Diffraction in incoherent light}
\author{Su-Heng Zhang$^1$, Shu Gan$^1$, De-Zhong Cao$^2$, Jun Xiong$^1$, Xiangdong
Zhang$^1$ and Kaige Wang$^1\footnote{%
Corresponding author: wangkg@bnu.edu.cn}$}
\address{1.Department of Physics, Applied Optics Beijing Area Major Laboratory,\\
Beijing Normal University, Beijing 100875, China\\
2.Department of Physics, Yantai University, Yantai 264005, China}

\begin{abstract}
Phase reversal occurs in the propagation of an electromagnetic wave
in a negatively refracting medium or a phase-conjugate interface.
Here we report the experimental observation of phase reversal
diffraction without the above devices. Our experimental results and
theoretical analysis demonstrate that phase reversal diffraction can
be formed through the first-order field correlation of chaotic
light. The experimental realization is similar to phase reversal
behavior in negatively refracting media.

\end{abstract}

\maketitle

Diffraction changes the wavefront of a travelling wave. A lens is a
key device that can modify the wavefront and perform imaging. The
complete recovery of the wavefront is possible if its phase evolves
backward in time in which case an object can be imaged to give an
exact copy. A phase-conjugate mirror formed by a four-wave mixing
process is able to generate the conjugate wave with respect to an
incident wave and thus achieve lensless imaging\cite {wolf}. A slab
of negative refractive-index material can play a role similar to a
lens in performing imaging\cite{ves,pen,smi,fang,dav,smo}. Pendry in
a recent paper\cite{pen2} explored the similarity between a
phase-conjugate interface and negative refraction, and pointed out
their intimate link to time reversal. An experiment to realize a
negatively refracting lens by means of phase-conjugate interfaces
was also proposed.

Here we report an experiment which demonstrates that phase reversal
diffraction can occur through the first-order spatial correlation of
chaotic light using neither negative refraction nor a
phase-conjugate interface. The experimental setup is an
interferometer as shown in Fig. 1. An object is placed in one arm of
the interferometer, while a glass rod with a refractive index
$n=1.5163$ is inserted in the other arm. The two ends of the rod are
plane. The interferometer is illuminated by an incoherent light
source, a Na lamp
of wavelength 589.3 nm with an extended illumination area of 10$\times $10 mm%
$^2$. Interference patterns of the interferometer can be recorded by
either of two CCD cameras. The travelling distances from the source
to the detection plane of the CCD
cameras through the object and reference arms are $z_o=$41.8 cm and $z_r=$%
33.8 cm, respectively. When the length of the glass rod is $l=15.5$
cm, the two arms of the interferometer have the same optical path
although their physical lengths are different. Under the
equal-optical-path condition, the two fields to be interfered in the
detection plane come from the same wavefront of the source. We will
show later that a medium in the path changes its diffraction length
away from optical path. The diffraction length of the reference path
can be calculated as $\overline{Z}=z_r-l+l/n=28.5$ cm. If phase
reversal diffraction exists, we may predict that the recovery of
wavefront will occur at a certain place. When the object is placed
at a distance $z_{o1}=\overline{Z}$ from the source, we observe its
image on the CCD screen, as shown in Fig. 2. The left column in Fig.
2 shows the images of an amplitude-modulated object of two Chinese
characters (China), and the right one is for a phase-modulated
object consisted of two transparent holes, having a path difference
of about a half wavelength. We can see that the two patterns of
Figs. 2(a) and 2(b), recorded by the two CCD cameras, respectively,
have a phase difference of $\pi $. Since BS$_2$ is a 50/50
beamsplitter, the intensity background can be eliminated through the
difference between (a) and (b), and the visibility of the image is
enhanced, as shown in Fig. 2(c). On the contrary, the images are
erased when we take the sum of (a) and (b), leaving only the
intensity background of the two beams, as shown in Fig. 2(d). For an
object consisting of the two holes, the images exhibit distinct
phase contrast: when one is bright, the other is dark. The imaging
scheme is evidently phase-sensitive.

We now use a double-slit of slit width $b=$125 $\mu $m and spacing $d=$ 300 $%
\mu $m as the object, and compare the interference patterns in the
same configuration using spatially incoherent and coherent light.
The double-slit is placed at the same position as above, and its
images are observed in the left part of Fig. 3. Then we insert a
pinhole in front of the lamp to improve the spatial coherence.
Instead of the image of the double-slit we observe its
interference fringes, as shown in the right part of Fig. 3, as
expected. As a matter of fact, for coherent interferometry,
lensless imaging can never occur no matter where the object is
placed. Next, for the incoherent light source, we move the
double-slit away from the position for imaging. Figures 4(a)-4(e)
show the
evolution of the interference patterns for $z_{o1}=31.0$, $28.5$, $24.2$, $%
20.0$, and $10.6$ cm, respectively, where 4(b) corresponds to the
image of the double-slit.

To understand the experiment, we consider optical diffraction
described by $E(x)=\int h(x,x^{\prime })E_s(x^{\prime })dx^{\prime
}$, where $E_s(x)$ and $E(x)$ are the field distributions before and
after the diffraction, respectively, and $h(x,x^{\prime })$ is the
impulse response function (IRF) of the transverse positions $x$ and
$x^{\prime }$. In the paraxial propagation, the IRF for free travel
over a distance $z$ in homogeneous material is given by\cite{good}
\begin{equation}
H(x,x_0;Z,\overline{Z})=\sqrt{k_0/(i2\pi \overline{Z})}\exp \left[ ik_0Z+ik_0(x-x_0)^2/(2%
\overline{Z})\right] ,  \label{1}
\end{equation}
where $k_0$ is the wave number in vacuum, and $Z=nz$ is the optical
path and $n$ is the refractive index. We define $\overline{Z}\equiv
z/n$ as the diffraction length in the medium.

In a successive diffraction through two media of lengths $l_j$ and
indices $n_j$ ($j$=1,2), we obtain the IRF of the system to be
\begin{equation}
H(x,x_0;n_1l_1+n_2l_2,l_1/n_1+l_2/n_2)=\int H(x,x^{\prime
};n_2l_2,l_2/n_2)H(x^{\prime },x_0;n_1l_1,l_1/n_1)dx^{\prime },
\label{6}
\end{equation}
where the optical path and the diffraction length in the resultant IRF are $%
n_1l_1+n_2l_2$ and $l_1/n_1+l_2/n_2$, respectively. In general, the
IRF for a series of cascaded media is still given by Eq. (\ref{1}),
where the optical path $Z=n_1l_1+n_2l_2+\cdots $ and the diffraction
length $\overline{Z}=l_1/n_1+l_2/n_2+\cdots $.

The diffraction length $\overline{Z}$ can be negative only when a
negatively refracting medium is present in the path. The diffraction
with a negative diffraction length is phase-reversed. In successive
diffraction through several media, the phase reversal diffraction in
a negatively refracting medium counteracts the normal one, which may
result in $\overline{Z}=0$. In this case the wavefront of the beam
is recovered exactly in the propagation, and lensless imaging
occurs. Hence, by means of the diffraction theory of light we have
illustrated the imaging condition $\overline{Z}=0$, which is
evidence for the presence of phase reversal diffraction.

To be specific, we consider the successive diffraction through two
media (see Fig. 5a). Let $T(x_0)$ describe a transmittance object,
illuminated by a plane wave $E_0$, then the outgoing field after
diffraction is written as
\begin{equation}
E(x)=E_0\int T(x_0)H(x,x_0;n_1l_1+n_2l_2,l_1/n_1+l_2/n_2)dx_0.
\label{4a}
\end{equation}
When $l_1/n_1+l_2/n_2=0$ is fulfilled, Eq. (\ref{6}) becomes
\begin{equation}
H(x,x_0;n_1l_1+n_2l_2,0)=\exp [ik_0(n_1l_1+n_2l_2)]\times \delta
(x-x_0). \label{4p}
\end{equation}
Hence we obtain the image $E(x)=E_0\exp [ik_0(n_1l_1+n_2l_2)]T(x)$,
and $l_1/n_1+l_2/n_2=0$ signifies the imaging condition. Otherwise,
Eq. (\ref{4a}) represents Fresnel diffraction with a diffraction
length of $l_1/n_1+l_2/n_2$.

We return to our experimental scheme in which the incoherent source
field $ E_s(x)$ is assumed to be quasi-monochromatic and satisfies
completely spatial incoherence $\langle E_s^{*}(x)E_s(x^{\prime
})\rangle =I_s\delta (x-x^{\prime })$, where $I_s$ is the intensity.
Let $E_o(x)$ and $E_r(x)$ be the field distributions of the object
and reference waves in the recording plane, respectively, then the
intensity pattern in the statistical average is given by $\langle
I(x)\rangle =\langle E_o^{*}(x)E_o(x)\rangle +\langle
E_r^{*}(x)E_r(x)\rangle +[\langle E_r^{*}(x)E_o(x)\rangle +$c.c.$].$
The interference term for the incoherent source is obtained as
\begin{equation}
\left\langle E_r^{*}(x)E_o(x)\right\rangle =I_s\int
h_r^{*}(x,x_0)h_o(x,x_0)dx_0,  \label{3}
\end{equation}
where $h_o(x,x_0)$ and $h_r(x,x_0)$ are the IRFs for the object and
reference arms, respectively. Equation (\ref{3}) establishes a joint
diffraction of two waves, one of which acts as a conjugate wave with
the other. As for the coherent light, however, the two fields are
separable and there is no joint diffraction between them.

In the interferometer, the IRF for the object arm is written as
\begin{equation}
h_o(x,x_0)=\int H(x,x^{\prime };z_{o2},z_{o2}) T(x^{\prime
})H(x^{\prime },x_0;z_{o1},z_{o1}) dx^{\prime }. \label{2}
\end{equation}
The IRF of the reference arm is described by Eq. (\ref{1}), i.e.
$h_r(x,x_0)=H(x,x_0;Z,\overline{Z})$. In the statistical correlation
of Eq. (\ref{3}), the reference wave acts as a conjugate wave that
reverses its phase,
$H^{*}(x,x_0;Z,\overline{Z})=H(x,x_0;-Z,-\overline{Z})$, in forming
a joint diffraction pattern with the object wave.  As shown in Fig.
5b, when the interferometer is opened out and the two arms are set
along a line, the joint diffraction through the two arms is
comparable with the successive diffraction through two media, one of
which is negatively refracting material.

To realize interference in the interferometer, the optical path
difference between the two arms must be less than the longitudinal
coherence length of the source. Hence we assume the equal optical
path lengths in the interferometer, i.e. $z_o=Z$. If the distance
between the object and the source is equal to the diffraction length
in the reference arm, $z_{o1}=\overline{Z},$ Eq. (\ref{3}) becomes
\begin{equation}
\left\langle E_r^{*}(x)E_o(x)\right\rangle =I_s\sqrt{k_0/(2\pi
iz_{o2})}T(x). \label{4}
\end{equation}
The object has been exactly reconstructed in the detection plane.
Otherwise, Eq. (\ref{3}) is written as
\begin{equation}
\left\langle E_r^{*}(x)E_o(x)\right\rangle = I_s \int T(x^{\prime
})H(x,x^{\prime };z_{o1}+z_{o2}-Z,Z_{eff}) dx^{\prime }, \label{5}
\end{equation}
where $Z_{eff}$ is the effective diffraction length, given by $%
1/Z_{eff}=1/z_{o2}+1/(z_{o1}-\overline{Z})$. Equation (\ref{5})
displays the Fresnel diffraction pattern of an object propagating a
distance $Z_{eff}$, which is composed of the two diffraction
lengths, $z_{o2}$ and $z_{o1}-\overline{Z}$. Both imaging equation
(\ref{4}) and diffraction equation (\ref{5}) can find their
counterparts in the negative refraction scheme (see Eq. (\ref{4a})).
We note that the intensities of the two arms, $\langle
E_o^{*}(x)E_o(x)\rangle $ and $\langle E_r^{*}(x)E_r(x)\rangle $,
are homogeneously distributed, contributing a flat background to the
interference pattern.

For the scheme of Fig. 1, the optical path of the reference arm is $%
Z=z_r-l+nl=41.8$ cm, which is identical to that of the object arm,
while the corresponding diffraction length
$\overline{Z}=z_r-l+l/n=28.5$ cm. The
latter determines the object's position for imaging, i.e. $z_{o1}|_{imaging}=%
\overline{Z}$, or equivalently, $z_{o2}|_{imaging}=Z-\overline{Z}%
=l(n-1/n)=13.3$ cm. When the refractive index is closer to unity,
the object's position for imaging approaches the detection plane.

Under the equal-optical-path condition, the effective diffraction
length can be expressed as
$Z_{eff}=z_{o2}[1-z_{o2}/z_{o2}|_{imaging}]$. Remarkably, our scheme
is capable of performing both normal diffraction and phase reversal
diffraction depending on whether the effective diffraction length is
positive or negative, respectively. A positive diffraction length is
obtained only when $z_{o2}<z_{o2}|_{imaging}$. Figure 4(a) shows the
near-field diffraction pattern for $Z_{eff}=2.0$ cm. When
$z_{o2}>z_{o2}|_{imaging}$, the diffraction patterns in Figs.
4(c)-4(e) correspond to the negative effective diffraction lengths
$Z_{eff}=-5.7,$ $-13.9,$ and $-42.0$ cm, respectively. To our
knowledge, this is the first experimental observation of phase
reversal diffraction pattern of an object.

In summary, we have demonstrated that phase reversal diffraction can
exist in an interferometer driven by incoherent light. When the
diffraction length in the reference arm is equal to the object
distance in the object arm, the wavefront of the object is
reconstructed in the outgoing plane due to the diffraction reversal
between the two arms. In the past few years, the phenomenon of
lensless ``ghost'' imaging with thermal light has attracted much
attention\cite {wkg,shih1,shih2,wla}; ghost imaging is performed
through intensity correlation measurements based on the
Hanbury-Brown and Twiss effect\cite {hbt,gigi}. Since the intensity
correlation measurement of thermal light records the modulus of the
first-order cross field correlation at two positions, i.e. $\left|
\langle E_r^{*}(x_1)E_o(x_2)\rangle \right| ^2$, the origin of
``lensless'' ghost imaging can be explained by reasoning similar to
the above. However, it should be pointed out that our interferometer
implements first-order interference, which is a basically different
phenomenon. The present scheme can be regarded as incoherent
interferometry: the first-order spatial interference in an
interferometer illuminated by incoherent light\cite{wkg1}. It thus
incorporates main properties of both coherent interference and
incoherent spatial field correlation. To realize the incoherent
interferometry, two arms of interferometer must undergo different
diffraction configurations. A positive refraction medium, inserted
in one arm of the interferometer, can replace the lens in the
interferometric scheme of Ref.\cite{wkg1}, just like a negative
refraction medium making a lens\cite{pen}. However, our imaging
scheme can bypass the problem of aberration that comes with lens
imaging. In comparison with similar effects in negative refraction
schemes, the evanescent wave cannot be recovered in the
interferometer containing only positively refracting materials,
hence the issue of surpassing the diffraction limit is still
unattainable in the present scheme. In practical applications with
ordinary optical devices, our scheme may provide a convenient
experimental platform for exploring phase reversal diffraction, and
may be valuable in the application of phase contrast imaging
techniques where coherent sources and lenses are unavailable, such
as when x-ray and electron beams are used.

The authors thank L. A. Wu for helpful discussions. This work was
supported by the National Fundamental Research Program of China,
Project No. 2006CB921404, and the National Natural Science
Foundation of China, Project No. 10874019.


\bigskip
Figure Caption:

Fig. 1 (Color on line) Experimental setup of the interferometer
formed by two mirrors, M$_1$ and M$_2$, and two beamsplitters,
BS$_1$ and BS$_2$. Object T is placed in one arm while a glass rod
is in the other. The interferometer is illuminated by a sodium lamp.

Fig. 2 Experimental results of lensless imaging of two
objects placed at the position of $z_{o1}|_{imaging}=28.5$ cm from the source (or $z_{o2}|_{imaging}=13.3$ cm from BS%
$_2$). Left column: for an amplitude-modulated object of two Chinese
characters (China); right column: for a phase-modulated object of
two holes with a phase difference of $\pi $. (a) and (b) are the
images recorded by the two CCDs; (c) is the difference of (a) and
(b), and (d) their sum, respectively.

Fig. 3 Experimental results of a double-slit object placed at the
same position as that in Fig. 2. Left column: image patterns when
the source is spatially incoherent; right column: fringe patterns
when the source is spatially coherent (with a pinhole aperture
behind the source). Both (a) and (b) were two patterns recorded by
the two CCDs.

Fig. 4 Diffraction patterns of the double-slit.
(a)through (e): the double-slit is placed, respectively, at the positions of $z_{o1}=31.0$, $28.5$, $24.2$, $20.0$, and $%
10.6$ cm, corresponding to the effective diffraction lengths
$Z_{eff}=2.0$, $0$, $-5.7$, $-13.9$, and $-42.0$ cm. We see that (b)
is the image of the double-slit.

Fig. 5 (Color on line) Illustration of lensless imaging by the two
schemes: (a) imaging via a successive diffraction through two media
where one is a negative refractive index medium; (b) imaging using
an incoherent light interferometer. When the interferometer is
opened out and the two arms are set along a line, the joint
diffraction through the two arms is comparable with that in (a).
\end{document}